\documentclass[prb,twocolumn,floats,aps,superscriptaddress]{revtex4}
\usepackage{graphicx,graphics,color,epsfig} 
\usepackage{multirow}
\usepackage{physics}

\newcommand{\vs}{\ensuremath{\mu_{s}}}
\newcommand{\vd}{\ensuremath{\mu_{d}}}
\newcommand{\vsd}{\ensuremath{V_{sd}}}
\newcommand{\ep}{\epsilon}

\begin{document}

\title{Seeing the strongly-correlated zero-bias anomaly in double quantum dot measurements}

\author{Rachel Wortis}
\affiliation{Department of Physics \& Astronomy, Trent University, Peterborough, Ontario, K9L0G2, Canada}
\author{Joshua Folk}
	\affiliation{Stewart Blusson Quantum Matter Institute, University of British Columbia, Vancouver, British Columbia, V6T1Z4, Canada}
	\affiliation{Department of Physics and Astronomy, University of British Columbia, Vancouver, British Columbia, V6T1Z1, Canada}
\author{Silvia L\"{u}scher}
	\affiliation{Stewart Blusson Quantum Matter Institute, University of British Columbia, Vancouver, British Columbia, V6T1Z4, Canada}
	\affiliation{Department of Physics and Astronomy, University of British Columbia, Vancouver, British Columbia, V6T1Z1, Canada}
\author{Sylvia Luyben}
\affiliation{Department of Physics \& Astronomy, Trent University, Peterborough, Ontario, K9L0G2, Canada}
\date{\today}

\begin{abstract}
Experiments in doped transition metal oxides often show suppression in the single-particle density of states at the Fermi level, but disorder-induced zero-bias anomalies in strongly-correlated systems remain poorly understood.
Numerical studies of the Anderson-Hubbard model have identified a zero-bias anomaly that is unique to strongly correlated materials, with a width proportional to the inter-site hopping amplitude $t$.\cite{Chiesa2008}
In ensembles of two-site systems, a zero-bias anomaly with the same parameter dependence also occurs, suggesting a similar physical origin.\cite{Wortis2010}
We describe how this kinetic-energy-driven zero-bias anomaly in ensembles of two-site systems may be seen in a mesoscopic realization based on double quantum dots.  
Moreover, the double-quantum-dot measurements provide access not only to the ensemble-average density of states but also to the details of the transitions which give rise to the zero-bias anomaly.
\end{abstract}


\maketitle

\section{Introduction}

Semiconductor-based quantum dots and transition-metal oxides are both subjects of current interest in condensed matter physics, yet with seemingly little connection to each other.  
Transition-metal oxides are studied typically as bulk materials in which atomic-scale Coulomb interactions profoundly shape the electronic structure and induce phenomena that range from high temperature superconductivity to colossal magnetoresistance.
Quantum dots, on the other hand, are typically built from semiconductors described by relatively simple band structures, with atomic-scale interactions folded into Fermi liquid parameters such as effective mass.
Here we show that a simple mesoscopic structure involving two coupled semiconductor-based quantum dots can provide unique insights into some of the behaviors of bulk disordered strongly-correlated materials.

Disordered interacting systems generically exhibit a feature in the density of states (DOS) pinned to the Fermi level, often referred to as a zero-bias anomaly (ZBA), the properties of which can be used as a probe of electronic structure.
Established theoretical frameworks exist for understanding the energy dependence of this feature in systems where the correlations are weak: In metals with weak interactions and disorder, a feature known as the Altshuler-Aronov zero-bias anomaly is found\cite{Altshuler1985}, and a corresponding feature in insulators is known as the Efros-Shklovskii Coulomb gap.\cite{Efros1975}
ZBAs are also observed when tunneling into transition metal oxides, but they generally do not match either the Altshuler-Aronov or the Efros-Shklovskii pictures.\cite{Sarma1998,Lahoud2014,Altshuler1985}

The Hubbard model, widely used in describing transition metal oxides, becomes the Anderson-Hubbard model under the addition of disorder in the site potentials.
Numerical studies of the Anderson-Hubbard model on two-dimensional lattices have established the existence of a ZBA with unique parameter dependence.\cite{Chiesa2008,Chen2012} A ZBA with the same parameter dependence is found in ensembles of two-site systems, providing physical insight into the origin of this feature.\cite{Wortis2010,Chen2010,Wortis2011} 

\begin{figure}
\includegraphics[width=\columnwidth]{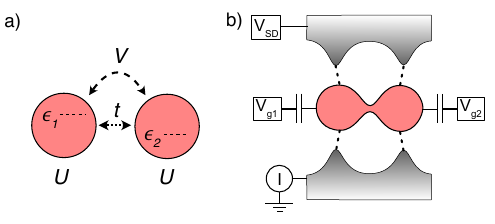}
\caption{\label{dqd}
(Color online.)  (a) Schematic diagram of a single two-site system.
(b) The DQD realization of the system in (a), with tunnel coupling $t$ between sites and interdot electrostatic coupling $V$.  The dots are connected to source and drain leads in the parallel configuration of relevance to this paper.
}
\end{figure}

This discovery opens the door to modeling the ZBA observed in disordered strongly-correlated materials using double quantum dot (DQD) structures, which have been commonplace for decades in the mesoscopics community.\cite{RevModPhys.75.1}.  DQDs can be described by a two-site Anderson-Hubbard model with each dot corresponding to a single site (Fig.\ \ref{dqd}).  The analog of disorder in a bulk system is variation of site energies in a DQD using independent gate voltages coupled to the two dots.  To researchers in bulk strongly-correlated materials, the message of this work is that DQDs provide a controlled environment in which to see the ZBA unique to disordered strongly-interacting systems.
To researchers in the semiconductor mesoscopics community, the message is that transport data through a simple DQD structure displays the physical origin of the strongly-correlated ZBA.

This paper begins with a brief review of the motivation for studying ensembles of two-site systems, a presentation of the two-site Anderson-Hubbard model and its realization in DQDs, and an overview of the ZBA in the ensemble-averaged DOS.  With this framework laid, we explore how this physics of the strongly-correlated ZBA is reflected in DQD measurements. 

\section{Motivation for studying the ensemble of two-site systems}
\label{motivation}

The motivation for looking at an ensemble of two-site systems was a set of surprising numerical results.  Quantum Monte Carlo and exact diagonalization studies of the Anderson-Hubbard model on two-dimensional lattices show a suppression in the single-particle density of states at the Fermi level with unique parameter dependence:\cite{Chiesa2008,Chen2012}
the ZBA width varies linearly with hopping amplitude and is independent of interaction strength, disorder strength, and filling, within specified ranges.
This contrasts with the results of mean-field calculations in which the ZBA depends on the strength of interactions\cite{Tusch1993,Fazileh2006,Chen2009} and of disorder.\cite{Shinaoka2010}
A physical picture for the mean-field ZBA centers on level repulsion:\cite{Levit1999,Chen2010}
While hopping in a tight-binding model always produces level repulsion, if it is irrespective of energy no ZBA is produced.
However, when a nonlocal Coulomb interaction is included and treated in a mean-field approximation, the exchange term of the interaction enhances the level repulsion specifically between states on opposite sides of the Fermi level, resulting in a ZBA.
Even when only onsite interactions $U$ are included in the model, an effective nonlocal interaction $V_{eff} \propto t^2/U$ may be generated through exchange terms.

\begin{figure}
\includegraphics[width=\columnwidth]{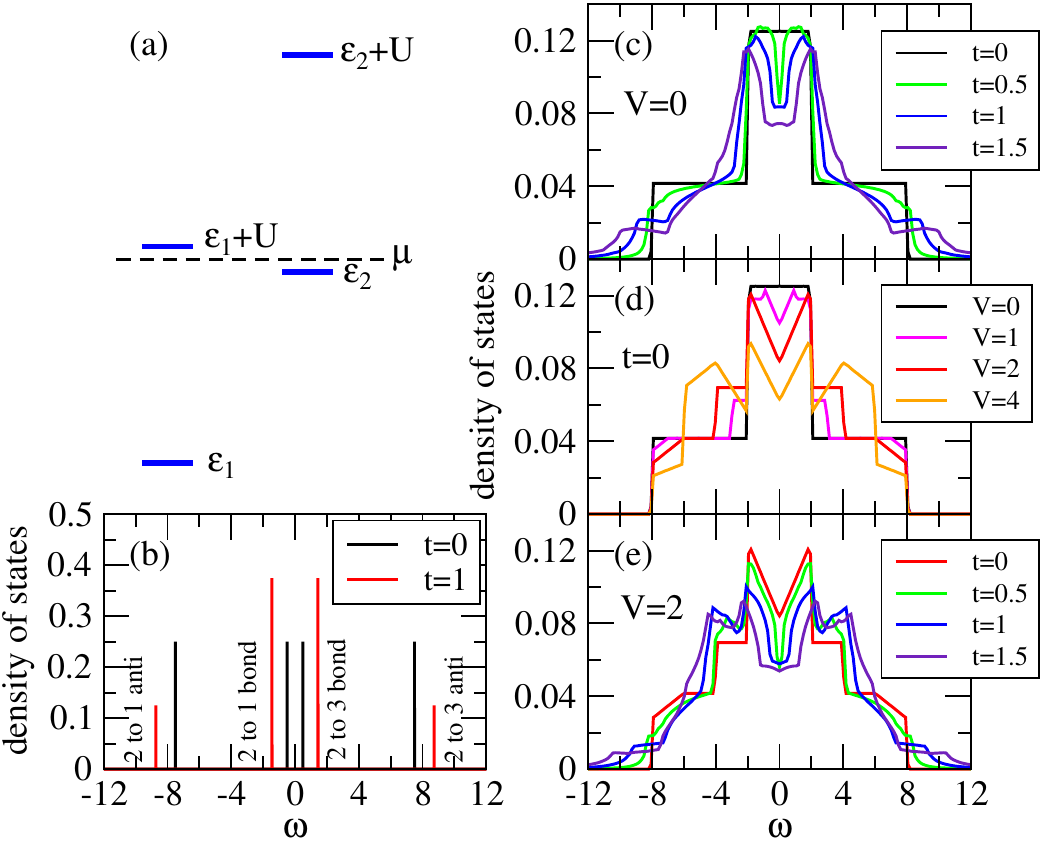}
\caption{\label{zba}
(Color online.)  (a) Diagram of one possible arrangement of Hubbard orbitals relative to the chemical potential with particular significance as discussed in the text.
(b) The DOS of the system shown in (a) without hopping (black) and with hopping (red).
(c) The ensemble-average DOS with $\Delta=12$, $U=8$, $V=0$ and $t$ as indicated.
(d) The ensemble average DOS with $\Delta=12$, $U=8$, $t=0$ and $V$ as indicated.
(e) The ensemble-average DOS with $\Delta=12$, $U=8$, $V=2$ and $t$ as indicated.
In all cases, $\mu=V+U/2$, corresponding to half filling.
}
\end{figure}

Looking for physical insight into these numerical results, one simplifying limit to consider is that of strong disorder.  For infinite disorder, the system is an ensemble of single sites, which has no ZBA.  One step beyond this is an ensemble of two-site systems, which does display a ZBA, and moreover one with the same parameter dependence as that found in lattices.\cite{Wortis2010,Chen2010,Wortis2011}
For the two-site ensemble, the linear increase in the spacing between the peaks at the edges of the anomaly is visible in Fig.\ \ref{zba}(c).
The lack of variation with other parameters, and within what limits, is explored in Ref.~\onlinecite{Wortis2010}.
The similarity in parameter dependence of the two-site ZBA with that found in larger systems suggests a shared physical origin and hence motivates further examination of the two-site case.
What is found is that the unique parameter dependence arises from a level alignment (Fig.\ \ref{zba}(a)) that occurs only in systems with both strong disorder and strong interactions, as discussed in detail in Section \ref{zba_in_ensavgdos} below.

While the primary focus of these studies has been the effect of strong onsite interactions in combination with disorder, we also consider nearest-neighbor interactions.
The added influence of nearest-neighbor interactions has been explored in bulk systems\cite{Song2009,Chen2012,Wortis2014} but was not included in prior discussion of ensembles of two-site systems.\cite{Wortis2010,Chen2010,Wortis2011} 
Nearest-neighbor interactions are necessarily present in DQD systems. 
In the absence of inter-site tunneling, the introduction of nearest-neighbor interactions in the ensemble of two-site systems results in a V-shaped minimum in the DOS at zero energy, Fig.\ \ref{zba}(d). 
Fig.~\ref{zba}(e) shows the evolution of the DOS with increasing tunneling for a fixed nonzero value of nearest-neighbor interaction strength.

\section{The two-site Anderson-Hubbard model and a double quantum dot}
\label{system}

We are considering systems consisting of two sites, each of which has a variable site potential and between which tunneling may occur (Fig.~\ref{dqd}a).
Such systems are described by the two-site Anderson-Hubbard model with the following Hamiltonian:
\begin{eqnarray}
{\hat H} &=& 
- t \sum_{\sigma=\uparrow, \downarrow} \left( {\hat c}_{1\sigma}^{\dag} {\hat c}_{2\sigma} + {\hat c}_{2\sigma}^{\dag} {\hat c}_{1\sigma} \right) \nonumber \\
& & + \sum_{i=1,2} \left( \epsilon_i {\hat n}_i + U {\hat n}_{i\uparrow} {\hat n}_{i\downarrow} \right)
\label{ahm}
\end{eqnarray}
${\hat c}_{i\sigma}$ and ${\hat n}_{i\sigma}$ are the annihilation and number operators for lattice site $i$ and spin $\sigma$.  
$t$ is the hopping amplitude and $U$ the strength of the on-site Coulomb repulsion.
Disorder is modelled by choosing $\epsilon_i$, the energy of the orbital at site $i$, from a uniform distribution of width $\Delta$.

DQDs offer an approach to realize this Hamiltonian in experiment (Fig.~\ref{dqd}b).
DQDs are commonly built in semiconductor-based two-dimensional electron gases using metal gates to confine and control the electrons\cite{RevModPhys.75.1}.  The energy of an orbital on dot 1, $\epsilon_1$, can be tuned by applying a voltage $V_{g1}$ to an adjacent gate.
Likewise for dot 2, $V_{g2}$ tunes $\epsilon_2$.  $U$ effectively represents  the single-dot charging energies $e^2/C$, where $C$ is the self capacitance of each dot, assumed to be the same. 
The height of the tunnel barrier between the two dots, and hence the hopping amplitude $t$, is typically controlled by an additional gate.
The two dots are generally close enough that the presence of an electron on dot 1 increases the potential at dot 2.  
In the tight-binding model (1) this electrostatic coupling can be represented by adding a nearest-neighbor interaction term 
$V{\hat n}_1 {\hat n}_2$ in the Hamiltonian.

Bulk tunneling measurements are approximated in DQD systems using a parallel configuration (Fig.~1b), in which both dots couple to both the source and the drain such that the tunneling process may occur through both dots simultaneously.\cite{Chan2002,holleitner2003,Chen2004,hubel2008,Hatano2013,keller2014}  This is in contrast to the more conventional series coupling in which electrons tunnel sequentially through the two dots.  

For the purpose of this presentation, we further specify characteristics of the coupling strength to the leads and of the bias between them.
First, the strength of the couplings to source and drain, characterized by rates $\gamma_S$ and $\gamma_d$ respectively, are assumed to be weak enough that tunneling to and from the reservoirs does not perturb the states in the dot; instead, current through the dot serves as a non-invasive probe of those levels.
In addition, the contact to the drain is much stronger than that to the source:  $\gamma_d \gg \gamma_s$, while both $\gamma_s$ and $\gamma_d$ remain in the small coupling regime.
Furthermore, the applied bias between source and drain $\vsd \equiv \vs-\vd$ is simultaneously larger than $k_B T$ while still being small enough that at most only a single transition falls between the source and the drain.  We define the source as the reservoir with higher chemical potential, such that electrons flow from source to drain.  
In this configuration, the DQD spends most of its time in its ground state as set by $\vd$, and current flows via the temporary occupation of excited states with one additional particle. 

In the proposed DQD realization described here, 
the size of each individual dot must be sufficiently small and hence the orbital level spacing sufficiently large
that we can assume the dots contain just 0, 1, or 2 electrons and higher orbitals may be neglected.
To observe the effects of interest here,
site potentials $\epsilon_1$ and $\epsilon_2$ are varied continuously over a range $\Delta>U$, large enough to cause the ground-state occupation of each dot to vary by $\pm 1$, while
$U$ must be large relative to the hopping amplitude $t$,
and the temperature must be less than all these energy scales.  The calculations shown here reflect $T=0$; the detailed lineshape would change but the ZBA would still be present for $k_B T$ nonzero so long as it is less than $t$.\cite{Wortis2011}.
We remind the reader that $\Delta$ in this paper refers to the disorder distribution width, not to orbital level spacing (as is common in DQD literature) as the orbital spacing is assumed to be large enough as not to play a role, that is, much larger than $k_B T$ for temperature $T$ and $eV_{sd}$ for source-drain bias $V_{sd}$.

\section{Origin of the zero-bias anomaly in the ensemble-average density of states}
\label{zba_in_ensavgdos}

Figures \ref{zba}(c)-(e) show the ensemble-averaged DOS of an ensemble of two-site Anderson-Hubbard systems in which both the onsite Coulomb repulsion $U$ and the disorder strength $\Delta$ are large, specifically $\Delta>U>>t$.  The opening of a ZBA as hopping (tunneling) is turned on is shown both without (panel (c)) and with (panel (e)) nearest-neighbor interactions.
Here, we describe the origin of this strongly-correlated ZBA starting first with a review of the spectrum of a single two-site system and then focusing on a particular level alignment that is accessible in these systems and gives rise to the ZBA.

In a single two-site system with no interactions ($U=V=0$), the DOS is just a histogram of the single-particle states as a function of energy.  Without hopping between the dots, these peaks are at the site energies $\ep_1$ and $\ep_2$, while with hopping they are at the energies of the bonding and anti-bonding states.
When interactions are included, the DOS is the density of single-particle transitions from the $N$-particle many-body ground state to $N\pm 1$-particle many-body excited states.
\begin{eqnarray}
\rho(\omega) &=& \frac{\rho_1(\omega) +\rho_2(\omega)}{2}
\end{eqnarray}
where $\rho_i(\omega)$ is the local DOS on dot $i$, given by the imaginary part of the local single-particle Green's function.
\begin{eqnarray}
\rho_i(\omega) &=& -\frac{1}{\pi} {\rm Im}\ G_{ii}(\omega)
\end{eqnarray}
We calculate the Green’s function using the exact eigenstates in the Lehmann representation.\cite{FetterWalecka}
The ground state is set by the chemical potential, corresponding to $\vd$ in the proposed experiment.
The number of available transitions depends on the ground state and the weight depends on the magnitude of the corresponding matrix elements.

A key point in the appearance of a linear-in-$t$ ZBA\cite{Wortis2010,Chen2010} is that when $\Delta>U$, level alignment $\ep_2=\ep_1\pm U$ is possible.  This is the scenario shown in Fig.\ \ref{zba}(a).
Furthermore, when $t/U<<1$, it is possible to have significant (linear in $t$) level repulsion between $\ep_2$ and $\ep_1+U$ while that between $\ep_2$ and $\ep_1$ is much smaller (of order $t^2/U$).
The presence of hopping in systems with this configuration results in a lowering of the energy of the 2-particle singlet state by an amount linear in $t$, a change which impacts the transitions that arise in an ensemble of two-site systems in three distinct ways:
First, for systems with a 2-particle singlet ground state, transitions to 1-particle and 3-particle excited states are increased in energy by an amount linear in $t$ because the energies of the 1- and 3-particle states are not changed to first order in $t$.  This shifts weight away from the Fermi level (Fig.\ \ref{zba}(b)).
Second, some systems that have 1-particle or 3-particle ground states when $t=0$, switch to a 2-particle ground state, hence modifying the available transitions.  
For later reference we will refer to these first two effects as the ground state effects.
And finally, for those systems that retain 1-particle and 3-particle ground states when hopping is turned on, transitions to 2-particle excited states are now split, with the transition to the singlet being lower in energy than that to the triplets.  For later reference we will refer to this as the splitting effect

The net result of these effects is the ZBA shown in Fig.\ \ref{zba}(c) and (e).  However, in this ensemble-average DOS the details of the effects are largely hidden.
In Section \ref{exp_sigs} we discuss how DQD measurements can show the effects explicitly, and how the ZBA in the ensemble-average DOS may be extracted from the DQD measurement. 

\section{Experimental signatures of the zero-bias anomaly}
\label{exp_sigs}

We discuss here two ways in which the physics of the ZBA is reflected in the DQD transport measurement sketched in Fig.\ \ref{dqd}(b).  We begin by focusing near the Fermi level, examining the changes hopping causes in the stability diagrams often produced in DQD measurements.  We then consider the dependence on the bias voltage between source and drain and connect the stability diagrams with the ensemble-average DOS.

\subsection{Seeing the reduction of the DOS near the Fermi level}
\label{exp_sigs_a}

\begin{figure}
\includegraphics[width=0.9\columnwidth]{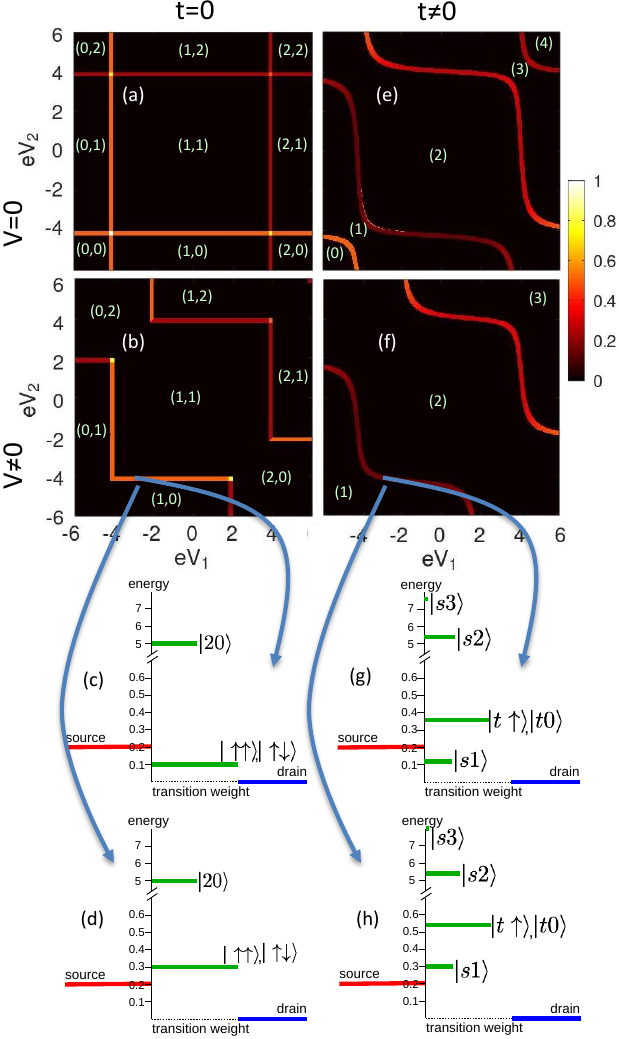}  
\caption{\label{levels+maps}(Color online.) 
Stability diagrams for a parallel-coupled double quantum dot system as described in the text with 
(a) no hopping, no electrostatic coupling;
(b) no hopping, $V=2$;
(e) $t=0.6$, $V=0$; and
(f) $t=0.6$, $V=2$.
The color scale indicates the magnitude of the current as a function of $V_{g1}$ and $V_{g2}$.
The ground-state occupation of each dot is indicated in the two $t=0$  plots, panels (a) and (b), while in panels (e) and (f) the numbers in parentheses indicate the overall ground-state occupation.
In all cases $\Delta=12$, $U=8$, $\mu=V+U/2$ and $\vsd=0.2$.
Transition spectrum diagrams for $(eV_1,eV_2)=(-3,-4.01)$ (c) $t=0$ and (g) $t\ne 0$;
and for $(eV_1,eV_2)=(-3,-4.03)$ (d) $t=0$ and (h) $t\ne 0$.
Diagrams showing $\vs$ on the left, $\vd$ on the right and the available single-particle-addition transitions.  The height of the transition lines indicates energy, and the length indicates the weight of the transition.  
}
\end{figure}

It is common practice to plot the current through a DQD system as a function of the gate voltages on the two dots, resulting in what is often called a stability diagram.  
This section focuses on a fixed, low value of $\vsd$, addressing how the resulting diagrams reflect the 
suppression near the Fermi level of the ensemble-average DOS shown in Fig.\ \ref{zba}. 
In the configuration detailed in Section \ref{system}, the stability diagram is a map of the DOS weight integrated over the range $0 \leq \omega \leq \vsd$.
Examples are shown in Fig.\ \ref{levels+maps} panels (a), (b), (e), and (f) for different settings of $t$ and $V$.

We note that the configuration described, specifically the stronger coupling to drain than source, introduces an asymmetry which has the benefit of allowing clear differentiation of the effects giving rise to the ZBA.  However, after taking an ensemble average, the ZBA itself will be symmetric about the Fermi level, as in a bulk system, and from an experimental point of view positive and negative bias will result in the same changes to net current.

To examine the connection between the DOS and the stability diagram, consider first the simplest case of two independent dots.  Current flows through one dot when a single-particle transition from the ground state has an energy less than $\vsd$.
These low energy transitions occur only for parameters very close to a change in the ground state occupation, so nonzero current flow appears in Fig.\ \ref{levels+maps}(a) only along lines marking the boundaries between different ground states.
For readers from the strongly correlated community, we note that we adopt the convention that the axes of the stability diagram correspond to increasing occupancy up and to the right.  
For readers from the mesoscopics community, first we emphasize that our stability diagrams are plotting current and not differential conductance. Second, we note that the distinct coloration of the lines at the $1\to 2$ particle and the $2 \to 3$ particle ground state transitions is a result of our choice to consider stronger coupling to the drain and $\vsd > k_B T$.  If the contact to the source was more open than that to the drain, or equivalently if in an experiment the sign of the bias is changed without changing the couplings, the current would flow through $N-1$ particle excited states, shifting the lines slightly up and to the right and reversing the coloration.  If $\vsd<k_B T$, the 1-to-2 and 2-to-3 lines would be the same colour, effectively an average of these two scenarios.  In all three cases the average of the current over the ensemble shows the same zero bias anomaly.

Next, consider the case of two parallel-coupled dots, nearby to each other and therefore not fully independent, but still without tunneling.  In this case, electrostatic coupling affects the ground state occupancy of each dot and hence the stability diagram, Fig.\ \ref{levels+maps}(b).  Fig.\ \ref{levels+maps}(c) shows an example of the transition spectrum (single-particle DOS) corresponding to a set of parameters for which  there is a transition with energy less than $\vsd$, so current does flow.  Indeed there are two such transitions, from the 1-particle ground state $\ket{\uparrow 0}$ to two different 2-particle excited states, $\ket{\uparrow \uparrow}$ and $\ket{\uparrow \downarrow}$, which are degenerate when $t=0$ because there is no singlet-triplet splitting.
In contrast, Fig.\ \ref{levels+maps}(d) shows an example of the transition spectrum for a system in which current does not flow:  Here the same transitions discussed above have an energy outside the window between source and drain.

Figure 2(c) of Ref.\ \onlinecite{Chan2002} shows an example of a measured stability diagram of the type shown in Fig.\ \ref{levels+maps}(b), with regions of fixed site occupancy separated by straight lines where current flows. 
Note that, while in a calculation each parameter may be tuned independently, in experiments they are often interdependent.  In particular, increasing $V_{g1}$ will also slightly increase the potential on dot 2, resulting in lines of current in the experimental diagram that are slanted rather than strictly horizontal and vertical lines as in the theory diagram.  This interdependence does not affect our result, and may be minimized in experiment by the use of virtual gates that are linear combinations of physical gate voltages.\cite{hensgens2017quantum}

Next, consider the case of non-zero $t$, when hopping is allowed between the two dots.  
The current through the system is proportional to the weight of transitions from the $N$-electron ground state to $N+1$-electron excited states that fall in the energy range $0<\omega<\vsd$.
Fig.\ \ref{levels+maps}(e) and (f) show examples of
the resulting stability diagram, without and with electrostatic coupling respectively.  Fig.\ \ref{levels+maps}(g) and (h) show examples of the transition spectra (single-particle DOS) for two different parameter settings that would and would not give rise to current flow, respectively.  
Experimental examples of the curved lines of current shown in Fig.\ \ref{levels+maps}(f) can be seen in Ref.\ \onlinecite{Chen2004} Fig.\ 1(b) and Ref.\ \onlinecite{Hatano2013} Fig.\ 1(b).

Comparing panels (a) and (e) of Fig.\ \ref{levels+maps}, we see that allowing tunneling between the dots changes both the shape and color of the lines.  
The shape change is due to changes in the ground state.  In particular, in the upper left and lower right (corresponding to $\ep_2 \sim \ep_1 \pm U$), there are more dot level settings which yield 2-particle ground states in (e) than (a), due to the ground state effects defined in Section \ref{zba_in_ensavgdos}.
The most striking color change is at the 1-to-2-particle transition.  The line becomes much darker with inter-dot tunneling because of the splitting effect defined in Section \ref{zba_in_ensavgdos}:  there are two possible transitions when $t=0$ (Fig.\ \ref{levels+maps}(c)) whereas when $t\ne 0$ there is only a single low-energy transition into the 2-particle singlet state (Fig.\ \ref{levels+maps}(g)).
The line at the 2-to-3-particle transition also changes color, becoming somewhat brighter, because the singlet ground state allows transitions into two possible 3-particle states whereas without inter-dot tunneling only a single spin option is available to a given Fock ground state.

The intent of this paper is to connect these DQD stability diagrams with the ZBA in the DOS of an ensemble of two-site systems shown in Fig.\ \ref{zba}(c)-(e).  
The experimental equivalent of the ensemble is to scan the gates $V_{g1}$ and $V_{g2}$ within an appropriate region of the stability diagram.  
The integrated current over such a scan then reflects the ensemble-average DOS.  In an experiment, tuning $t$ is likely to effect the coupling to source and drain, making a quantitative comparison of these ensemble-average results challenging.  Nonetheless, the stability diagrams themselves display the transport signatures that lead to the reduction of integrated current with hopping $t$ and nearest-neighbor interaction $V$.

A comparison of Figs.\ \ref{levels+maps} panels (a) and (b) shows the effect of $V$ alone, in the absence of hopping:  The outward shift of the lines of current reduces the ensemble-average DOS at the Fermi level as shown in Fig.\ \ref{zba}(d), but the transitions involved don't change in character.
In contrast, the addition of tunneling $t$ changes both the shape and the intensity of the lines of current, as shown in Fig.\ \ref{levels+maps} (a) and (e) for $V=0$ and in (b) and (f) for $V \ne 0$.
The combined effect of these changes is a net decrease in the ensemble-averaged DOS at the Fermi level shown for $V=0$ in Fig.\ \ref{zba} (c) and for $V \ne 0$ in Fig.\ \ref{zba} (e): 
The increase in weight at the 2-to-3-particle transition is more than canceled by the decrease in weight at the 1-to-2-particle transition, and the decrease in the length of these lines due to their curvature further suppresses the average.

\begin{figure}[htp]
\includegraphics[width=0.9\columnwidth]{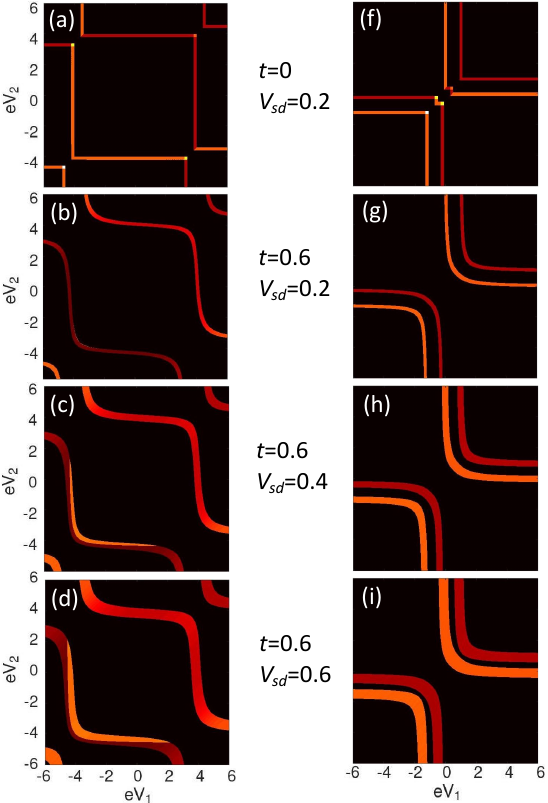} \\ \vskip 0.1 in
\includegraphics[width=0.8\columnwidth]{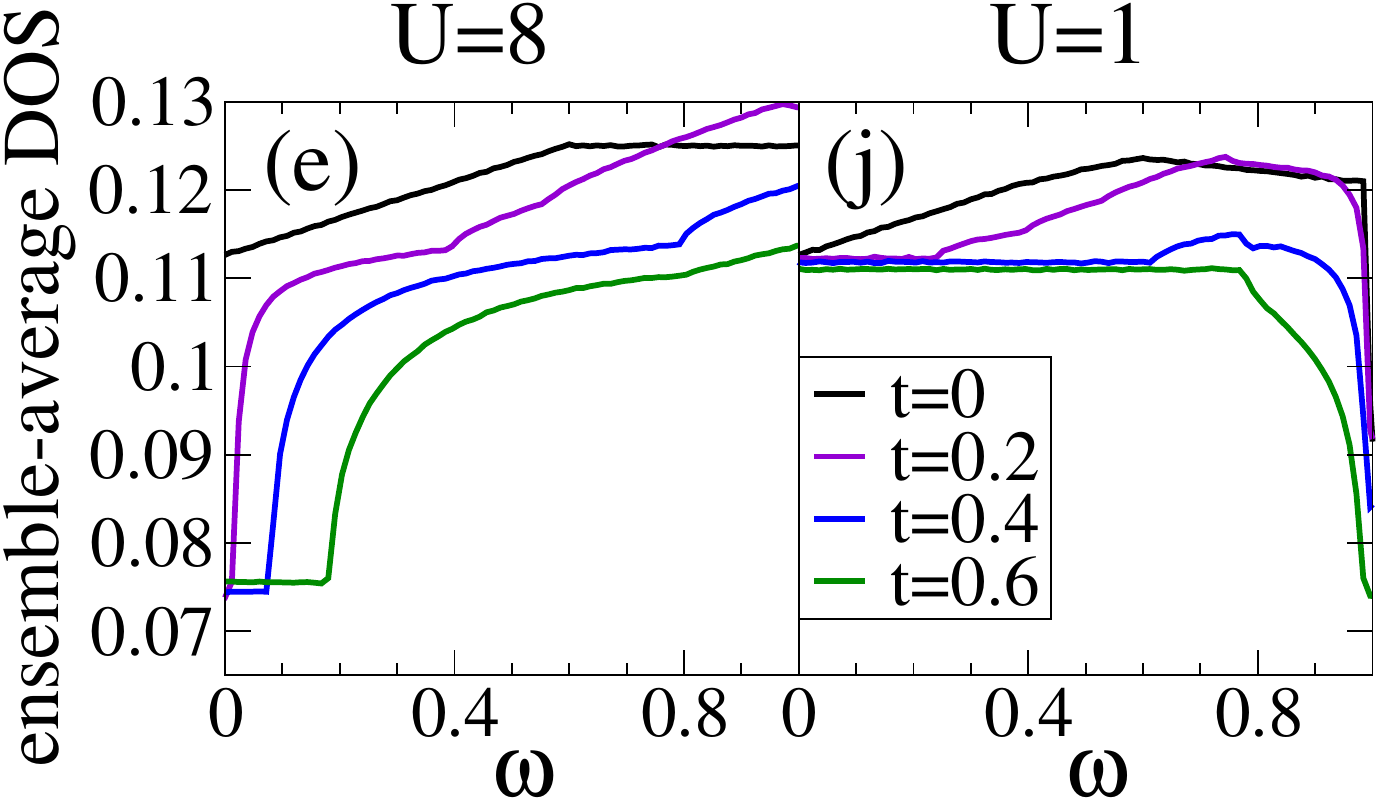}
\caption{(Color online.) 
Stability diagrams for (a)-(d) $U=8$ and 
(f)-(i) $U=1$ with $t$ and $\vsd$ as indicated and the same color scale as shown in Fig.\ \ref{levels+maps}. 
The ensemble-average DOS as a function of frequency near the Fermi level for (e) $U=8$ and (j) $U=1$. In all cases, $\Delta=12$, $V=0.6$, $\mu=V+U/2$.}
\label{Vbias}
\end{figure}

\subsection{Seeing the energy dependence of the zero-bias anomaly}
\label{exp_sigs_b}

The DOS at the Fermi level can be suppressed for multiple reasons, and, if only the ensemble-average value is accessible, these can be difficult to distinguish.  Stability diagrams provide additional information allowing physics unique to strong correlations to be disentangled from other effects.  Further details appear when the dependence on the bias voltage is considered, specifically in the form of $I(\vsd)$

Figure \ref{Vbias} panels (a)-(d) and (f)-(i), like those in Fig.\ \ref{levels+maps}, show the weight of the DOS in the range $0 \leq \omega \leq \vsd$, but this time for several values of $\vsd$.
Experimentally, these correspond to stability diagrams representing the d.c.\ current measured over a range of $V_{g1},V_{g2}$, for fixed $\vsd$.
Again, we consider a system in which the coupling to the drain is significantly more open than that to the source and $\vsd > k_B T$ such that the DQD is primarily in its ground state, with current flowing via brief occupation of single-particle-addition excited states.

Figures \ref{Vbias} (b)-(d) show the evolution of the stability diagram with increasing $\vsd$ for a strongly correlated system $t/U\ll1$.  At the lowest bias, the 1-to-2-particle transition exhibits sharply reduced current due to the splitting effect defined in \ref{zba_in_ensavgdos} and discussed further in \ref{exp_sigs_a}.  As the bias is increased, all the lines of current broaden, and at the 1-to-2-particle transition a bright orange sliver appears, corresponding to the entrance of the triplet transitions into the window between source and drain.  This occurs first in the region of $\ep_2 \sim \ep_1$ where the splitting is lower in magnitude, and extends outward to the $\ep_2 \sim \ep_1 \pm U$ regions.

Figure\ \ref{Vbias} (e) shows the ensemble-average DOS for the same interaction strengths $U$ and $V$ and a range of $t$ values.
The $t=0$ curve (black) corresponds to Fig.\ \ref{Vbias}(a) and shows the ensemble-average DOS suppression associated with nearest-neighbor Coulomb repulsion $V$ as shown in Fig.\ \ref{zba}(d).
The $t=0.6$ curve (green) corresponds to Figs.\ \ref{Vbias}(b)-(d).
The simple broadening of the lines is reflected in the flat DOS at very low energy, 
while the entrance of the triplet transitions is seen in the sharp rise in the DOS at $\omega \approx 0.2$

A natural question to consider is what is special about the case of strong correlations.
As a point of comparison, Figs.\ \ref{Vbias} (f)-(j) show the same quantities with $U=1$ such that $t/U \sim 1$ (shown for illustration, although implementing $U\ll\Delta$ in a DQD system is not feasible).
All the physics just discussed is still present at very small $t$ values in a small region at the center of the diagram.  However,  
when $t \sim U$, it is large enough to couple not only $\ep_2$ with $\ep_1$, or only $\ep_2$ with $\ep_1+U$, but instead to couple all of these states together. As a result, the stability diagrams lose the unique features associated with changes in the transition spectrum and the ensemble-average DOS shows no ZBA.
Comments on the comparison with mean-field treatments can be found in Ref.\ \onlinecite{Chen2010}.

\section{Conclusion}

In summary, we have described how DQD measurements can show, explicitly, the changes in the transition spectrum that result in a ZBA in ensembles of two-site systems.  This ZBA arises exclusively when onsite interactions are strong, and it has been shown to share the same parameter dependence as that found in numerical studies of two-dimensional lattices.
In bulk systems measurements have limited control over parameters and can access only the average DOS, making it difficult to differentiate between different mechanisms of DOS suppression.
For example, the simple linear-$t$ dependence of the width of the ZBA in Fig.\ \ref{zba}(c) is obscured by the presence of nearest neighbor interaction $V$.
DQD measurements by contrast offer exceptional tunability and produce stability diagrams which readily display the physical origins of DOS features.
In particular, both the gapping of the triplet state and its re-emergence at higher bias is clearly visible in the stability diagrams both without and with $V$.
Observing this effect in DQD systems would establish a first solid point of contact between theory and experiment in the study of the disorder-driven ZBA in strongly correlated systems.

\section{Acknowledgements}
R.W. gratefully acknowledges support from the Natural Sciences and Engineering Research Council (NSERC) of Canada and thanks the Pacific Institute for Theoretical Physics for hosting her.  J.F. and S.L. acknowledge support from NSERC, the Canada Foundation for Innovation, the Canadian Institute for Advanced Research, and the Stewart Blusson Quantum Matter Institute.


\end{document}